\documentclass[aip,reprint,twocolumn,amsmath,amssymb,groupedaddress
]{revtex4-1}

\usepackage{graphicx}
\usepackage{dcolumn}
\usepackage{bm}

\begin{document}

\title{Spin-current diode with a monoaxial chiral magnet}

\author{Shun~Okumura$^1$, Hiroaki~Ishizuka$^1$, Yasuyuki~Kato$^1$, Jun-ichiro~Ohe$^2$, and Yukitoshi~Motome$^1$}

\affiliation{$^1$Department of Applied Physics, The University of Tokyo, Tokyo 113-8656, Japan\\
$^2$Department of Physics, Toho University, Chiba 274-8510, Japan }

\begin{abstract}
Monoaxial chiral magnets exhibit a chiral conical magnetic state in a magnetic field parallel to the chiral axis. 
The conical spins carry the potential for nonreciprocal transport phenomena, as they break both spatial inversion and time reversal symmetries. 
Here we study the spin-dependent transport in the chiral conical magnetic state, using the Landauer method based on Green's functions for a one-dimensional Kondo lattice model.
We show that the system exhibits nonreciprocal spin transport, which depends on the chirality, period, cone angle, and the polarization of the spin current.
In particular, we find the distinct cone angle dependence between the spin textures with long and short periods. 
We also show that the nonreciprocity is related with the spin states of itinerant electrons near the leads.
Our results indicate that the chiral cone acts as a spin-current diode, which can be flexibly controlled by a magnetic field.
\end{abstract}

\maketitle

Nonreciprocal transport has recently attracted considerable attention in materials science and its applications.
Although reciprocal relation is a fundamental principle in thermodynamics, it can be violated by breaking certain symmetries of the system.
A typical example is the $p$-$n$ junction, which exhibits a nonreciprocal electric current.
Such diode effects are also studied in bulk materials without spatial inversion and time reversal symmetries~\cite{Rikken2001}. 

As a candidate for the nonreciprocal transport, chiral magnets have attracted much interest since they break both of spatial inversion and time reversal symmetries.
The chiral magnetic structures are often stabilized by an antisymmetric exchange interaction, called the Dzyaloshinskii-Moriya interaction~\cite{Dzyaloshinskii1958,Moriya1960}, which originates from the lack of spatial inversion symmetry in the crystalline structure.
Chiral magnetic conductors show unconventional transport, such as a topological Hall effect in a skyrmion crystal~\cite{Lee2009,Neubauer2009,Schulz2012} and nonlinear negative magnetoresistance in a chiral soliton lattice~\cite{Togawa2013}.
Furthermore, a nonreciprocal electric current, which can be switched by the chirality and the magnetic field direction, has been reported and dubbed as the electrical magnetochiral effect~\cite{Rikken2001,Rikken2005,Yokouchi2017,Aoki2019}. 

\begin{figure}[t]
\centering
\includegraphics[width=\columnwidth,clip]{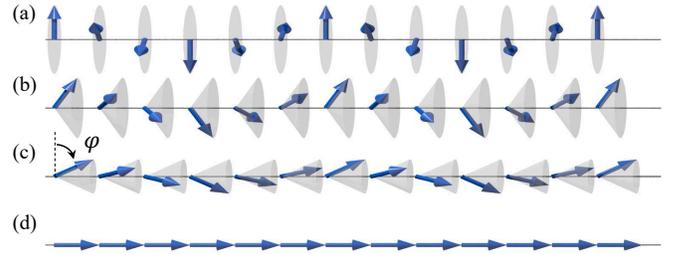}
\caption{Schematic pictures of (a) a chiral helimagnetic state (CHM), (b) and (c) chiral conical magnetic states (CCMs), (d) a forced ferromagnetic  state (FFM). The chiral axis is taken along the chain direction. The gray circles and cones are guides to the eye. $\varphi$ in (c) represents the cone angle. 
}
\label{f1}
\end{figure}

A monoaxial chiral magnetic conductor CrNb$_3$S$_6$ has been studied since the pioneering works by Dzyaloshinskii in 1960s~\cite{Dzyaloshinskii1964,Dzyaloshinskii1965}. 
This compound exhibits a chiral helimagnetic state (CHM) with the spatial period of $\sim40$ Cr sites at zero field~[Fig.~\ref{f1}(a)]~\cite{Togawa2012} and turns into a chiral conical magnetic state (CCM) under the magnetic field parallel to the chiral axis~[Figs.~\ref{f1}(b) and \ref{f1}(c)].
As the magnetic field increases, the cones are closed and finally relaxed into a forced ferromagnetic state (FFM) above the critical magnetic field [Fig.~\ref{f1}(d)]~\cite{Miyadai1983}. 
More recently, in YbNi$_3$Al$_9$, the CHM with a much shorter magnetic period ($\sim 3.75$ Yb sites at zero field) has been found~\cite{Miyazaki2012,Ohara2014}.
This compound also exhibits qualitatively the same CCM under the magnetic field parallel to the chiral axis~\cite{Yamashita2011}.
Meanwhile, when the magnetic field is applied perpendicular to the chiral axis, these monoaxial chiral magnets exhibit another chiral spin structure called the chiral soliton lattice~\cite{Kishine2005,Togawa2012,Togawa2016,Matsumura2017,Ninomiya2018}.
These peculiar magnetic properties have been theoretically studied by using spin-only models in which the itinerant electron degrees of freedom are integrated out~\cite{Nishikawa2016, Shinozaki2016,Laliena2016-2,Masaki2017,Laliena2017}. 
In addition, peculiar electrical transport and electronic states have been investigated~\cite{Kishine2011,Okumura2017,Okumura2018JPSJ,Okumura2018PhysB,Tokushuku2017, Watanabe2016}.
However, the spin transport is less studied in both experiment and theory.
In particular, nonreciprocal spin transport, which is interesting from both fundamental and application viewpoints, has not been fully understood for not only the CHM but also the CCM in an applied magnetic field. 

In this letter, we theoretically study nonreciprocal spin transport in the monoaxial chiral magnets. 
By using the Landauer method based on Green's functions, we calculate the spin-dependent conductance of itinerant electrons coupled with the chiral magnetic textures.
We show that a nonreciprocal spin current can be generated in both CHM and CCM, by an interfacial effect near the leads on the ballistic electrons.
The nonreciprocity depends on the chirality, period, cone angle, and the polarization of the spin current. 
Our results indicate that the chiral magnetic conductors are useful as a spin-current diode.

\begin{figure}[b]
\centering
\includegraphics[width=\columnwidth,clip]{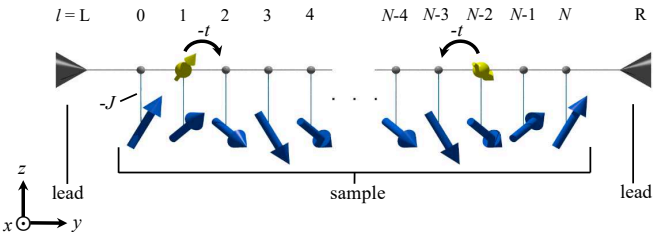}
\caption{Schematic figure of an ($N$+1)-site chain (sample) coupled to the leads at both ends. Itinerant electrons (yellow spheres with arrows) are coupled to localized spins (blue arrows) at each site.
}
\label{f2}
\end{figure}

We consider a one-dimensional chain connected to leads at both ends, as shown in Fig.~\ref{f2}. 
The one-dimensional chain, which we call ``sample'', is described by the Kondo lattice Hamiltonian: 
\begin{align}
	\mathcal{H}_\mathrm{K} = -t\sum_{l, \sigma}(c^{\dagger}_{l\sigma}c^{\;}_{l+1\sigma}+\mathrm{h.c.})-J\sum_{l, \sigma,\sigma'}c^{\dagger}_{l\sigma}{\boldsymbol \sigma}_{\sigma\sigma'}c^{\;}_{l\sigma'}\cdot{\mathbf S}_l,
	\label{eq:H_K}
\end{align}
where $c_{l\sigma}$ ($c^{\dagger}_{l\sigma}$) is the annihilation (creation) operator of the itinerant electron with spin $\sigma=\uparrow,\downarrow$ at site $l=0,\cdots, N$, $\boldsymbol{\sigma} = (\sigma^x,\sigma^y,\sigma^z)$ is the vector of Pauli matrices, and ${\mathbf S}_{l}=(S_l^x,S_l^y,S_l^z)$ represents the localized spin at site $l$. 
Here we take the $y$ axis along the chain direction as shown in Fig.~\ref{f2}.
The first term in Eq.~(\ref{eq:H_K}) describes the kinetic energy of itinerant electrons; the sum is taken for the nearest-neighbor sites from $l=0$ to $N-1$.
Hereafter we set the hopping energy scale as $t=1$.
The second term is for the onsite coupling between the itinerant electrons and the localized spins; the sum is taken from $l=0$ to $N$. 
We treat the localized spins as classical vectors for simplicity. 
In the following, we take $J=2$; we confirm that the results remain qualitatively the same for $J\neq0$~\cite{SM}.

We assume the localized spin state as $(S_l^x,S_l^y,S_l^z)=(\sin{Ql}\cos\varphi,\sin\varphi,\cos{Ql}\cos\varphi)$, where $Q$ is the wave number and $\varphi$ represents the cone angle (see Fig.~\ref{f1}).
The spin configuration describes the CCM [Figs.~\ref{f1}(b) and \ref{f1}(c)], including the CHM in the $xz$ plane at $\varphi=0$ [Fig.~\ref{f1}(a)] and the FFM along the $y$ axis at $\varphi=\pm90^{^\circ}$ [Fig.~\ref{f1}(d)].
In the following calculations, we fix the number of sites in the sample at $N+1=241$ (we confirm that the finite-size effect is negligibly small) and choose $Q$ in the range of $-\pi\leq Q\leq\pi$ so that $QN$ becomes a multiple of $2\pi$ (commensurate); in this setting, the localized spins at both ends $l=0$ and $N$ point to the same direction [${\bf S}_0 = {\bf S}_N = (0,\sin \varphi, \cos \varphi)$].
We define the chirality of the CCM as left- and right-handed for $-\pi<Q<0$ and $0<Q<\pi$, respectively.
  
This Kondo lattice chain is coupled to the leads which are described by noninteracting electron systems; the total Hamiltonian reads
\begin{align}
	\mathcal{H} &= \mathcal{H}_\mathrm{K} + \sum_{k}\left(\varepsilon_{\mathrm{L}k}d^\dagger_{\mathrm{L}k}d^{\;}_{\mathrm{L}k} + \varepsilon_{\mathrm{R}k}d^\dagger_{\mathrm{R}k}d^{\;}_{\mathrm{R}k}\right)\nonumber\\
	&+ \sum_{k,\sigma}\left(V_{\mathrm{L}k,0\sigma}d^\dagger_{\mathrm{L}k}c^{\;}_{0\sigma} + V_{\mathrm{R}k,N\sigma}d^\dagger_{\mathrm{R}k}c^{\;}_{N\sigma} + \mathrm{h.c.}\right),
\end{align}
where $d_{\mathrm{L(R)}k}$ is the annihilation operator for the $k$th eigenstate in the left (right) lead, $\varepsilon_{\mathrm{L(R)}k}$ is the eigenenergy for the $k$th state in the left (right) lead, and $V_{\mathrm{L(R)}k,0(N)\sigma}$ is the hybridization between the $k$th state in the left (right) lead and site 0 ($N$) in the sample. 

We investigate the nonreciprocal spin transport in this one-dimensional system by using the spin-dependent Landauer method~\cite{Ando1991,Ohe2007,Watanabe2016}. 
In this approach, by assuming a ballistic motion of electrons, the hybridization between the sample and the leads are taken into account as the self-energy given by
\begin{align}
	\Sigma_{lm}^{\sigma\sigma'}(\mu) = -\frac{i}{2}\left[\Gamma_\mathrm{L}^\sigma(\mu)\delta_{l\mathrm{L}}\delta_{m\mathrm{L}} + \Gamma_\mathrm{R}^\sigma(\mu)\delta_{l\mathrm{R}}\delta_{m\mathrm{R}}\right]\delta_{\sigma\sigma'},
	\label{eq:Sigma}
\end{align}
where $\mu$ is the chemical potential and $\Gamma_\mathrm{L(R)}^\sigma(\mu) = 2\pi\sum_{k}|V_{\mathrm{L(R)}k,0(N)\sigma}|^2\delta(\mu-\varepsilon_{\mathrm{L(R)}k})$; for simplicity, we employ the wide-band limit approximation~\cite{Jauho1994} and assume $\Gamma_\mathrm{L(R)}^\sigma(\mu)$ to be a constant $\Gamma=1$.
In the following calculations, we take $\mu=-0.5$ (the results remain qualitatively the same for arbitrary $\mu\neq 0$, with a difference in the sign for $\mu>0$). 
In order to avoid spurious oscillations of Green's function from the finite-size effect, we average the results by randomly sampling $\mu$ in the range of $-0.6<\mu<-0.4$. 
We take $10^3$ samples to estimate the statistical errors.

By using the self-energy in Eq.~(\ref{eq:Sigma}), retarded Green's function is obtained from Dyson's equation as 
\begin{align}
	\hat{\mathcal{G}}(\mu) = \left[\hat{\mathcal{H}}_\mathrm{K}-\hat{\Sigma}(\mu)-\mu\hat{I}\right]^{-1},
\end{align}
where $\hat{\mathcal{H}}_{\mathrm{K}}$, $\hat{\Sigma}(\mu)$, and $\hat{I}$ are the $(2N+6)\times(2N+6)$ matrix forms of the Kondo lattice Hamiltonian, self-energy, and identity matrix, respectively.
By using $\hat{\mathcal{G}}$, we calculate the difference between the conductance of right- and left-going spin currents as 
\begin{align}
\Delta G^\lambda = G_0\mathrm{Tr}\left[\mathcal{G}^\dagger_{\mathrm{L},\mathrm{R}}\sigma^\lambda\mathcal{G}^{\;}_{\mathrm{L},\mathrm{R}} - \mathcal{G}^\dagger_{\mathrm{R},\mathrm{L}}\sigma^\lambda\mathcal{G}^{\;}_{\mathrm{R},\mathrm{L}}\right], 
\end{align} 
where $\lambda$ is the polarization axis of the spin current ($\lambda=x,y,z$), and $\mathcal{G}_{\mathrm{L},\mathrm{R}}$ and $\mathcal{G}_{\mathrm{R},\mathrm{L}}$ are off-diagonal components of $\hat{\mathcal{G}}$, which are represented as the matrices in 2$\times$2 spin space.
Here, $G_0$ is the conductance quantum, $G_0=2e^2/h\sim7.75\times10^{-5}~\mathrm{\Omega}^{-1}$, where $e$ is the elementary charge and $h$ is the Planck constant.

\begin{figure}[t]
\centering
\includegraphics[width=0.85\columnwidth,clip]{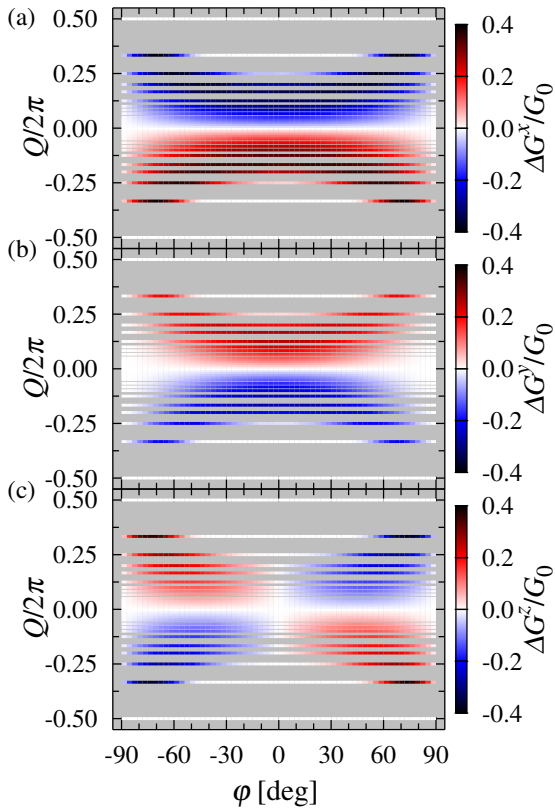}
\caption{
Differences between the conductance of right- and left-going spin currents, $\Delta G^{\lambda}$, as functions of $\varphi$ and $Q$: (a), (b), and (c) for the $x$, $y$, and $z$ polarized spin currents, respectively.
The gray region represents for incommensurate $Q$.
}
\label{f3}
\end{figure}

Figure~\ref{f3} shows the $\varphi$ and $Q$ dependence of $\Delta G^\lambda$. 
We find that all the components of $\Delta G^\lambda$ become nonzero when the system breaks the inversion symmetry ($Q \neq 0$, $\pm\pi$, nor $\varphi\neq\pm 90^\circ$), and they change the sign according to the chirality (sign of $Q$).
On the other hand, the $\varphi$ dependence depends on $\lambda$; $\Delta G^x$ and $\Delta G^y$ are even functions of $\varphi$ while $\Delta G^z$ is odd.
In addition, the $\varphi$ dependence is qualitatively different for small $|Q|$ (long periods) and large $|Q|$ (short).

\begin{figure}[b]
\centering
\includegraphics[width=0.85\columnwidth,clip]{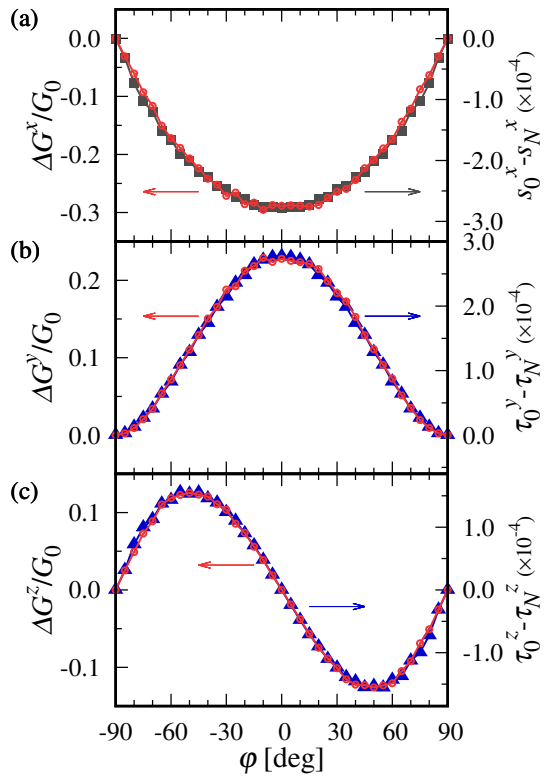}
\caption{
$\varphi$ dependence of $\Delta G^{\lambda}$ for (a) $x$, (b) $y$, and (c) $z$ polarized spin currents denoted by open circles.
The filled squares and triangles represent the differences of (a) the expectation values of the itinerant-electron spins~[Eq.~(\ref{eq:s})], and (b)(c) the magnetic torque for the itinerant-electron spins~[Eq.~(\ref{eq:t})] between the left- and right-ends of the sample, respectively.
Numerical errors are within the size of symbols.
We set the ordering vector $Q=2\pi/10$ (right-handed chirality with a long period).
}
\label{f4}
\end{figure}

Let us first take a closer look at the results for longer periods.
Figure~\ref{f4} shows the results at $Q=2\pi/10$.
As shown in Figs.~\ref{f4}(a) and \ref{f4}(b), $\Delta G^x$ and $\Delta G^y$ become nonzero except for $\varphi=\pm 90^\circ$ (FFM).
This indicates that both spin currents polarized along the $x$ and $y$ axes exhibit nonreciprocal behavior for the CCM as well as the CHM. 
The nonreciprocity takes the maximum for the CHM at $\varphi=0$ and decreases monotonically to zero as $\varphi \to \pm90^\circ$. 
On the other hand, as shown in Fig.~\ref{f4}(c), the spin current polarized along the $z$ axis is reciprocal for the CHM at $\varphi=0$, but it becomes nonreciprocal and chirality dependent for the CCM with $0<|\varphi|<90^\circ$.  
In contrast to Figs.~\ref{f4}(a) and \ref{f4}(b), the nonreciprocity takes the maximum at $|\varphi|\simeq45^\circ$.

In order to clarify the origin of the $\varphi$ dependences of $\Delta G^\lambda$, we focus on the spin states of the itinerant electrons near the leads. 
We here consider the difference of the itinerant electron spins at both ends, which is defined by
\begin{align}
	s^\lambda_0 - s^\lambda_N=\frac{1}{2}\sum_{\epsilon_j<\mu}\langle j|\sigma^\lambda_{0}|j\rangle - \frac{1}{2}\sum_{\epsilon_j<\mu}\langle j|\sigma^\lambda_{N}|j\rangle,
	\label{eq:s}
\end{align}
where $\epsilon_j$ and $|j\rangle$ are the $j$th eigenvalue and eigenvector of $\hat{\mathcal{H}}_{\mathrm{K}}$, respectively, and $\sigma^\lambda_l/2$ represents the $\lambda$ component of a spin operator at site $l$ (we set $\hbar = h/2\pi = 1$).
This quantity can be nonzero since the itinerant electron spins cannot follow the directions of the localized spins exactly and accumulate the deviations through the transport.
We find that only the $x$ component becomes nonzero in our setup.
This is because the projection of the localized spin configuration onto the $y$ and $z$ axes are symmetric with respect to the center of the system, while it is antisymmetric for $x$.
As shown in Fig.~\ref{f4}(a), the $\varphi$ dependence of $\Delta G^x$ well scales to that of $s_0^x-s_N^x$; both of them approximately follow $\propto\cos\varphi$.
The results suggest that the nonzero $\Delta G^x$ can be interpreted as a magnetoresistance predominantly caused by the relative angle between the end spins. 

In contrast, similar scaling does not work for the $y$ and $z$ spin currents. 
Instead, here we calculate the difference of the magnetic torque for itinerant-electron spins from localized moments given by
\begin{align}
	\tau^\lambda_0 - \tau^\lambda_N =\left[\mathbf{s}_0\times\mathbf{S}_0\right]^\lambda - \left[\mathbf{s}_N\times\mathbf{S}_N\right]^\lambda,
	\label{eq:t}
\end{align} 
where $\bold{s}_l = (s^x_l, s^y_l, s^z_l)$ is calculated as in Eq.~(\ref{eq:s}).
Due to the same reason discussed above, the $x$ component, $\tau^x_0-\tau^x_N$, vanishes but the $y$ and $z$ components become nonzero.
We find that the $\varphi$ dependences well scale to those of the nonreciprocity as shown in Figs.~\ref{f4}(b) and \ref{f4}(c); $\propto\cos^2\varphi$ for $\lambda=y$ and $\propto\sin2\varphi$ for $\lambda=z$.
The results suggest that the magnetic torque by the internal magnetic field from localized spins near the leads plays a role in the nonreciprocity in these two components.

\begin{figure}[t]
\centering
\includegraphics[width=0.85\columnwidth,clip]{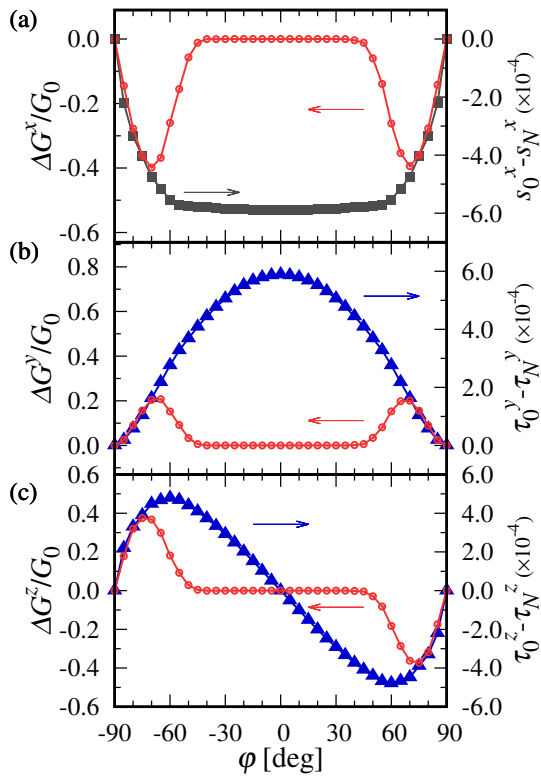}
\caption{
$\varphi$ dependence of $\Delta G^{\lambda}$ and the related magnetic properties of the itinerant electrons for $Q=2\pi/3$ (right-handed chirality with a short period).
}
\label{f5}
\end{figure}

Next, we discuss the cases with shorter periods.
Figure~\ref{f5} presents the results of $\Delta G^\lambda$ at $Q=2\pi/3$.
In stark contrast to the long period cases, $\Delta G^\lambda$ deviates from the related edge quantities and are strongly suppressed for small $|\varphi|$. 
This is presumably because the itinerant-electron spins cannot follow the rapid twist of localized spins with the short period.
As $|\varphi|$ increases, the relative angle between neighboring localized spins becomes smaller and the itinerant-electron spins can follow the localized spins again; hence the scaling behaviors are recovered for large $|\varphi|$
\footnote{We confirm this picture by examining the dependence on the strength of the spin-charge coupling $J$~\cite{SM}}.  

Our analysis on the simple model indicates that the nonreciprocity reaches the maximum values $|\Delta G^\lambda|\sim 3\times10^{-5}~\mathrm{\Omega}^{-1}$.
When we assume the length of the sample $\sim0.1~\mu\mathrm{m}$ along the $y$ axis, the maximum spin conductivity can be $\sim10^2~\Omega^{-1}\mathrm{m}^{-1}$, which is comparable to the previous study in noncentrosymmetric spin-orbit coupled systems~\cite{Hamamoto2017}.
In experiment, thin films of CrNb$_3$S$_6$ and YbNi$_3$Al$_9$ have been reported, whose thickness are $0.1~\mu\mathrm{m}\sim28~\mathrm{nm}$~\cite{Togawa2015,Wang2017} and $0.5~\mu\mathrm{m}$~\cite{Aoki2018} along the $c$ axis, respectively.
Our results suggest that such thin films can be used as a spin-current diode in spintronics.
Inversely, the present effect would be useful to detect CCMs through the nonreciprocity. 
While we assumed the CCM texture intact through the transport, effects of the spin transfer torque from spin-polarized electric current are also worth investigating for the dynamics of the chiral spin textures~\cite{Bostrem2008-1,Bostrem2008-2,Bostrem2009,Kishine2010}.

In the analyses above, we have assumed commensurate $Q$, which results in the different behavior between the $x$ ad $z$ components due to the direction of the end spins.
In generic cases including incommensurate $Q$ (gray region in Fig.~\ref{f3}), our results suggest that the nonreciprocity for spin currents with the spin polarization perpendicular to the chiral axis depends on the relative angle between the localized spins at both ends, but one parallel to the chiral axis does not.
In such a case, both spin polarization and magnetic torque contribute to the nonreciprocal spin current polarized perpendicular to the chiral axis.

In summary, we investigated nonreciprocal spin transport in a one-dimensional Kondo lattice model with a chiral conical magnetic order. 
By using the spin-dependent Landauer approach, we found nonreciprocal behaviors of the spin current which depend on the polarization direction of the injected spin current. 
Our detailed analyses revealed a relation of the current to the itinerant-electron spin polarization and magnetic torque near the edges.
We showed that this nonreciprocal spin current can be controlled by the magnetic field, chirality, and period of the spin textures.
In particular, we find the distinct cone angle dependence between the chiral spin textures with long and short periods.
Our finding indicates that monoaxial chiral magnets are potentially useful for spintronics as a spin-current diode.\\

This research was supported by JST CREST (JPMJCR18T2). 
This work was also supported by the Chirality Research Center in Hiroshima University and JSPS Core-to-Core Program, Advanced Research Networks.
S. O. is supported by JSPS through a research fellowship for young scientists.
J.O. was supported by CREST, JST, and a Grant-inAid for Scientific Research (A) (Grant No.17H01052) from MEXT, Japan.

\bibliography{ref.bib}

\providecommand{\noopsort}[1]{}\providecommand{\singleletter}[1]{#1}%
\begin{thebibliography}{45}%
\makeatletter
\providecommand \@ifxundefined [1]{%
 \@ifx{#1\undefined}
}%
\providecommand \@ifnum [1]{%
 \ifnum #1\expandafter \@firstoftwo
 \else \expandafter \@secondoftwo
 \fi
}%
\providecommand \@ifx [1]{%
 \ifx #1\expandafter \@firstoftwo
 \else \expandafter \@secondoftwo
 \fi
}%
\providecommand \natexlab [1]{#1}%
\providecommand \enquote  [1]{``#1''}%
\providecommand \bibnamefont  [1]{#1}%
\providecommand \bibfnamefont [1]{#1}%
\providecommand \citenamefont [1]{#1}%
\providecommand \href@noop [0]{\@secondoftwo}%
\providecommand \href [0]{\begingroup \@sanitize@url \@href}%
\providecommand \@href[1]{\@@startlink{#1}\@@href}%
\providecommand \@@href[1]{\endgroup#1\@@endlink}%
\providecommand \@sanitize@url [0]{\catcode `\\12\catcode `\$12\catcode
  `\&12\catcode `\#12\catcode `\^12\catcode `\_12\catcode `\%12\relax}%
\providecommand \@@startlink[1]{}%
\providecommand \@@endlink[0]{}%
\providecommand \url  [0]{\begingroup\@sanitize@url \@url }%
\providecommand \@url [1]{\endgroup\@href {#1}{\urlprefix }}%
\providecommand \urlprefix  [0]{URL }%
\providecommand \Eprint [0]{\href }%
\providecommand \doibase [0]{http://dx.doi.org/}%
\providecommand \selectlanguage [0]{\@gobble}%
\providecommand \bibinfo  [0]{\@secondoftwo}%
\providecommand \bibfield  [0]{\@secondoftwo}%
\providecommand \translation [1]{[#1]}%
\providecommand \BibitemOpen [0]{}%
\providecommand \bibitemStop [0]{}%
\providecommand \bibitemNoStop [0]{.\EOS\space}%
\providecommand \EOS [0]{\spacefactor3000\relax}%
\providecommand \BibitemShut  [1]{\csname bibitem#1\endcsname}%
\let\auto@bib@innerbib\@empty
\bibitem [{\citenamefont {Rikken}, \citenamefont {F{\"o}lling},\ and\
  \citenamefont {Wyder}(2001)}]{Rikken2001}%
  \BibitemOpen
  \bibfield  {author} {\bibinfo {author} {\bibfnamefont {G.~L. J.~A.}\
  \bibnamefont {Rikken}}, \bibinfo {author} {\bibfnamefont {J.}~\bibnamefont
  {F{\"o}lling}}, \ and\ \bibinfo {author} {\bibfnamefont {P.}~\bibnamefont
  {Wyder}},\ }\href@noop {} {\bibfield  {journal} {\bibinfo  {journal} {Phys.
  Rev. Lett.}\ }\textbf {\bibinfo {volume} {87}},\ \bibinfo {pages} {23602}
  (\bibinfo {year} {2001})}\BibitemShut {NoStop}%
\bibitem [{\citenamefont {Dzyaloshinsky}(1958)}]{Dzyaloshinskii1958}%
  \BibitemOpen
  \bibfield  {author} {\bibinfo {author} {\bibfnamefont {I.}~\bibnamefont
  {Dzyaloshinsky}},\ }\href@noop {} {\bibfield  {journal} {\bibinfo  {journal}
  {J. Phys. Chem. Solids}\ }\textbf {\bibinfo {volume} {4}},\ \bibinfo {pages}
  {241} (\bibinfo {year} {1958})}\BibitemShut {NoStop}%
\bibitem [{\citenamefont {Moriya}(1960)}]{Moriya1960}%
  \BibitemOpen
  \bibfield  {author} {\bibinfo {author} {\bibfnamefont {T.}~\bibnamefont
  {Moriya}},\ }\href@noop {} {\bibfield  {journal} {\bibinfo  {journal} {Phys.
  Rev.}\ }\textbf {\bibinfo {volume} {120}},\ \bibinfo {pages} {91} (\bibinfo
  {year} {1960})}\BibitemShut {NoStop}%
\bibitem [{\citenamefont {Lee}\ \emph {et~al.}(2009)\citenamefont {Lee},
  \citenamefont {Kang}, \citenamefont {Onose}, \citenamefont {Tokura},\ and\
  \citenamefont {Ong}}]{Lee2009}%
  \BibitemOpen
  \bibfield  {author} {\bibinfo {author} {\bibfnamefont {M.}~\bibnamefont
  {Lee}}, \bibinfo {author} {\bibfnamefont {W.}~\bibnamefont {Kang}}, \bibinfo
  {author} {\bibfnamefont {Y.}~\bibnamefont {Onose}}, \bibinfo {author}
  {\bibfnamefont {Y.}~\bibnamefont {Tokura}}, \ and\ \bibinfo {author}
  {\bibfnamefont {N.~P.}\ \bibnamefont {Ong}},\ }\href@noop {} {\bibfield
  {journal} {\bibinfo  {journal} {Phys. Rev. Lett.}\ }\textbf {\bibinfo
  {volume} {102}},\ \bibinfo {pages} {186601} (\bibinfo {year}
  {2009})}\BibitemShut {NoStop}%
\bibitem [{\citenamefont {Neubauer}\ \emph {et~al.}(2009)\citenamefont
  {Neubauer}, \citenamefont {Pfleiderer}, \citenamefont {Binz}, \citenamefont
  {Rosch}, \citenamefont {Ritz}, \citenamefont {Niklowitz},\ and\ \citenamefont
  {B{\"o}ni}}]{Neubauer2009}%
  \BibitemOpen
  \bibfield  {author} {\bibinfo {author} {\bibfnamefont {A.}~\bibnamefont
  {Neubauer}}, \bibinfo {author} {\bibfnamefont {C.}~\bibnamefont
  {Pfleiderer}}, \bibinfo {author} {\bibfnamefont {B.}~\bibnamefont {Binz}},
  \bibinfo {author} {\bibfnamefont {A.}~\bibnamefont {Rosch}}, \bibinfo
  {author} {\bibfnamefont {R.}~\bibnamefont {Ritz}}, \bibinfo {author}
  {\bibfnamefont {P.~G.}\ \bibnamefont {Niklowitz}}, \ and\ \bibinfo {author}
  {\bibfnamefont {P.}~\bibnamefont {B{\"o}ni}},\ }\href@noop {} {\bibfield
  {journal} {\bibinfo  {journal} {Phys. Rev. Lett.}\ }\textbf {\bibinfo
  {volume} {102}},\ \bibinfo {pages} {186601} (\bibinfo {year}
  {2009})}\BibitemShut {NoStop}%
\bibitem [{\citenamefont {Schulz}\ \emph {et~al.}(2009)\citenamefont {Schulz},
  \citenamefont {Ritz}, \citenamefont {Bauer}, \citenamefont {Halder},
  \citenamefont {Wagner}, \citenamefont {Franz}, \citenamefont {Pfleiderer},
  \citenamefont {Everschor}, \citenamefont {Garst},\ and\ \citenamefont
  {Rosch}}]{Schulz2012}%
  \BibitemOpen
  \bibfield  {author} {\bibinfo {author} {\bibfnamefont {T.}~\bibnamefont
  {Schulz}}, \bibinfo {author} {\bibfnamefont {R.}~\bibnamefont {Ritz}},
  \bibinfo {author} {\bibfnamefont {A.}~\bibnamefont {Bauer}}, \bibinfo
  {author} {\bibfnamefont {M.}~\bibnamefont {Halder}}, \bibinfo {author}
  {\bibfnamefont {M.}~\bibnamefont {Wagner}}, \bibinfo {author} {\bibfnamefont
  {C.}~\bibnamefont {Franz}}, \bibinfo {author} {\bibfnamefont
  {C.}~\bibnamefont {Pfleiderer}}, \bibinfo {author} {\bibfnamefont
  {K.}~\bibnamefont {Everschor}}, \bibinfo {author} {\bibfnamefont
  {M.}~\bibnamefont {Garst}}, \ and\ \bibinfo {author} {\bibfnamefont
  {A.}~\bibnamefont {Rosch}},\ }\href@noop {} {\bibfield  {journal} {\bibinfo
  {journal} {Phys. Rev. Lett.}\ }\textbf {\bibinfo {volume} {102}},\ \bibinfo
  {pages} {186601} (\bibinfo {year} {2009})}\BibitemShut {NoStop}%
\bibitem [{\citenamefont {Togawa}\ \emph {et~al.}(2013)\citenamefont {Togawa},
  \citenamefont {Kousaka}, \citenamefont {Nishihara}, \citenamefont {Inoue},
  \citenamefont {Akimitsu}, \citenamefont {Ovchinnikov},\ and\ \citenamefont
  {Kishine}}]{Togawa2013}%
  \BibitemOpen
  \bibfield  {author} {\bibinfo {author} {\bibfnamefont {Y.}~\bibnamefont
  {Togawa}}, \bibinfo {author} {\bibfnamefont {Y.}~\bibnamefont {Kousaka}},
  \bibinfo {author} {\bibfnamefont {S.}~\bibnamefont {Nishihara}}, \bibinfo
  {author} {\bibfnamefont {K.}~\bibnamefont {Inoue}}, \bibinfo {author}
  {\bibfnamefont {J.}~\bibnamefont {Akimitsu}}, \bibinfo {author}
  {\bibfnamefont {A.~S.}\ \bibnamefont {Ovchinnikov}}, \ and\ \bibinfo {author}
  {\bibfnamefont {J.}~\bibnamefont {Kishine}},\ }\href@noop {} {\bibfield
  {journal} {\bibinfo  {journal} {Phys. Rev. Lett.}\ }\textbf {\bibinfo
  {volume} {111}},\ \bibinfo {pages} {197204} (\bibinfo {year}
  {2013})}\BibitemShut {NoStop}%
\bibitem [{\citenamefont {Rikken}\ and\ \citenamefont
  {Wyder}(2005)}]{Rikken2005}%
  \BibitemOpen
  \bibfield  {author} {\bibinfo {author} {\bibfnamefont {G.~L. J.~A.}\
  \bibnamefont {Rikken}}\ and\ \bibinfo {author} {\bibfnamefont
  {P.}~\bibnamefont {Wyder}},\ }\href@noop {} {\bibfield  {journal} {\bibinfo
  {journal} {Phys. Rev. Lett.}\ }\textbf {\bibinfo {volume} {94}},\ \bibinfo
  {pages} {016601} (\bibinfo {year} {2005})}\BibitemShut {NoStop}%
\bibitem [{\citenamefont {Yokouchi}\ \emph {et~al.}(2017)\citenamefont
  {Yokouchi}, \citenamefont {Kanazawa}, \citenamefont {Kikkawa}, \citenamefont
  {Morikawa}, \citenamefont {Shibata}, \citenamefont {Arima}, \citenamefont
  {Taguchi}, \citenamefont {Kagawa},\ and\ \citenamefont
  {Tokura}}]{Yokouchi2017}%
  \BibitemOpen
  \bibfield  {author} {\bibinfo {author} {\bibfnamefont {T.}~\bibnamefont
  {Yokouchi}}, \bibinfo {author} {\bibfnamefont {N.}~\bibnamefont {Kanazawa}},
  \bibinfo {author} {\bibfnamefont {A.}~\bibnamefont {Kikkawa}}, \bibinfo
  {author} {\bibfnamefont {D.}~\bibnamefont {Morikawa}}, \bibinfo {author}
  {\bibfnamefont {K.}~\bibnamefont {Shibata}}, \bibinfo {author} {\bibfnamefont
  {T.}~\bibnamefont {Arima}}, \bibinfo {author} {\bibfnamefont
  {Y.}~\bibnamefont {Taguchi}}, \bibinfo {author} {\bibfnamefont
  {F.}~\bibnamefont {Kagawa}}, \ and\ \bibinfo {author} {\bibfnamefont
  {Y.}~\bibnamefont {Tokura}},\ }\href@noop {} {\bibfield  {journal} {\bibinfo
  {journal} {Nat. Commun.}\ }\textbf {\bibinfo {volume} {8}},\ \bibinfo {pages}
  {866} (\bibinfo {year} {2017})}\BibitemShut {NoStop}%
\bibitem [{\citenamefont {Aoki}, \citenamefont {Kousaka},\ and\ \citenamefont
  {Togawa}(2019)}]{Aoki2019}%
  \BibitemOpen
  \bibfield  {author} {\bibinfo {author} {\bibfnamefont {R.}~\bibnamefont
  {Aoki}}, \bibinfo {author} {\bibfnamefont {Y.}~\bibnamefont {Kousaka}}, \
  and\ \bibinfo {author} {\bibfnamefont {Y.}~\bibnamefont {Togawa}},\
  }\href@noop {} {\bibfield  {journal} {\bibinfo  {journal} {Phys. Rev. Lett.}\
  }\textbf {\bibinfo {volume} {122}},\ \bibinfo {pages} {057206} (\bibinfo
  {year} {2019})}\BibitemShut {NoStop}%
\bibitem [{\citenamefont {Dzyaloshinskii}(1964)}]{Dzyaloshinskii1964}%
  \BibitemOpen
  \bibfield  {author} {\bibinfo {author} {\bibfnamefont {I.~E.}\ \bibnamefont
  {Dzyaloshinskii}},\ }\href@noop {} {\bibfield  {journal} {\bibinfo  {journal}
  {Sov. Phys.--JETP}\ }\textbf {\bibinfo {volume} {19}},\ \bibinfo {pages}
  {960} (\bibinfo {year} {1964})}\BibitemShut {NoStop}%
\bibitem [{\citenamefont {Dzyaloshinskii}(1965)}]{Dzyaloshinskii1965}%
  \BibitemOpen
  \bibfield  {author} {\bibinfo {author} {\bibfnamefont {I.~E.}\ \bibnamefont
  {Dzyaloshinskii}},\ }\href@noop {} {\bibfield  {journal} {\bibinfo  {journal}
  {Sov. Phys.--JETP}\ }\textbf {\bibinfo {volume} {20}},\ \bibinfo {pages}
  {223} (\bibinfo {year} {1965})}\BibitemShut {NoStop}%
\bibitem [{\citenamefont {Togawa}\ \emph {et~al.}(2012)\citenamefont {Togawa},
  \citenamefont {Koyama}, \citenamefont {Takayanagi}, \citenamefont {Mori},
  \citenamefont {Kousaka}, \citenamefont {Akimitsu}, \citenamefont {Nishihara},
  \citenamefont {Inoue}, \citenamefont {Ovchinnikov},\ and\ \citenamefont
  {Kishine}}]{Togawa2012}%
  \BibitemOpen
  \bibfield  {author} {\bibinfo {author} {\bibfnamefont {Y.}~\bibnamefont
  {Togawa}}, \bibinfo {author} {\bibfnamefont {T.}~\bibnamefont {Koyama}},
  \bibinfo {author} {\bibfnamefont {T.}~\bibnamefont {Takayanagi}}, \bibinfo
  {author} {\bibfnamefont {S.}~\bibnamefont {Mori}}, \bibinfo {author}
  {\bibfnamefont {Y.}~\bibnamefont {Kousaka}}, \bibinfo {author} {\bibfnamefont
  {J.}~\bibnamefont {Akimitsu}}, \bibinfo {author} {\bibfnamefont
  {S.}~\bibnamefont {Nishihara}}, \bibinfo {author} {\bibfnamefont
  {K.}~\bibnamefont {Inoue}}, \bibinfo {author} {\bibfnamefont {A.~S.}\
  \bibnamefont {Ovchinnikov}}, \ and\ \bibinfo {author} {\bibfnamefont
  {J.}~\bibnamefont {Kishine}},\ }\href@noop {} {\bibfield  {journal} {\bibinfo
   {journal} {Phys. Rev. Lett.}\ }\textbf {\bibinfo {volume} {108}},\ \bibinfo
  {pages} {107202} (\bibinfo {year} {2012})}\BibitemShut {NoStop}%
\bibitem [{\citenamefont {Miyadai}\ \emph {et~al.}(1983)\citenamefont
  {Miyadai}, \citenamefont {Kikuchi}, \citenamefont {Kondo}, \citenamefont
  {Sakka}, \citenamefont {Arai},\ and\ \citenamefont {Ishikawa}}]{Miyadai1983}%
  \BibitemOpen
  \bibfield  {author} {\bibinfo {author} {\bibfnamefont {T.}~\bibnamefont
  {Miyadai}}, \bibinfo {author} {\bibfnamefont {K.}~\bibnamefont {Kikuchi}},
  \bibinfo {author} {\bibfnamefont {H.}~\bibnamefont {Kondo}}, \bibinfo
  {author} {\bibfnamefont {S.}~\bibnamefont {Sakka}}, \bibinfo {author}
  {\bibfnamefont {M.}~\bibnamefont {Arai}}, \ and\ \bibinfo {author}
  {\bibfnamefont {Y.}~\bibnamefont {Ishikawa}},\ }\href@noop {} {\bibfield
  {journal} {\bibinfo  {journal} {J. Phys. Soc. Japan}\ }\textbf {\bibinfo
  {volume} {52}},\ \bibinfo {pages} {1394} (\bibinfo {year}
  {1983})}\BibitemShut {NoStop}%
\bibitem [{\citenamefont {Miyazaki}\ \emph {et~al.}(2012)\citenamefont
  {Miyazaki}, \citenamefont {Aoki}, \citenamefont {Higashinaka},\ and\
  \citenamefont {Sano}}]{Miyazaki2012}%
  \BibitemOpen
  \bibfield  {author} {\bibinfo {author} {\bibfnamefont {R.}~\bibnamefont
  {Miyazaki}}, \bibinfo {author} {\bibfnamefont {Y.}~\bibnamefont {Aoki}},
  \bibinfo {author} {\bibfnamefont {R.}~\bibnamefont {Higashinaka}}, \ and\
  \bibinfo {author} {\bibfnamefont {H.}~\bibnamefont {Sano}},\ }\href@noop {}
  {\bibfield  {journal} {\bibinfo  {journal} {Phys. Rev. B}\ }\textbf {\bibinfo
  {volume} {86}},\ \bibinfo {pages} {155106} (\bibinfo {year}
  {2012})}\BibitemShut {NoStop}%
\bibitem [{\citenamefont {Ohara}\ \emph {et~al.}(2014)\citenamefont {Ohara},
  \citenamefont {Fukuta}, \citenamefont {Ohta}, \citenamefont {Kono},
  \citenamefont {Yamashita}, \citenamefont {Matsumoto},\ and\ \citenamefont
  {Yamaura}}]{Ohara2014}%
  \BibitemOpen
  \bibfield  {author} {\bibinfo {author} {\bibfnamefont {S.}~\bibnamefont
  {Ohara}}, \bibinfo {author} {\bibfnamefont {S.}~\bibnamefont {Fukuta}},
  \bibinfo {author} {\bibfnamefont {K.}~\bibnamefont {Ohta}}, \bibinfo {author}
  {\bibfnamefont {H.}~\bibnamefont {Kono}}, \bibinfo {author} {\bibfnamefont
  {T.}~\bibnamefont {Yamashita}}, \bibinfo {author} {\bibfnamefont
  {Y.}~\bibnamefont {Matsumoto}}, \ and\ \bibinfo {author} {\bibfnamefont
  {J.}~\bibnamefont {Yamaura}},\ }\href@noop {} {\bibfield  {journal} {\bibinfo
   {journal} {JPS Conf. Proc.}\ }\textbf {\bibinfo {volume} {3}},\ \bibinfo
  {pages} {017016} (\bibinfo {year} {2014})}\BibitemShut {NoStop}%
\bibitem [{\citenamefont {Yamashita}, \citenamefont {Ohara},\ and\
  \citenamefont {Sakamoto}(2011)}]{Yamashita2011}%
  \BibitemOpen
  \bibfield  {author} {\bibinfo {author} {\bibfnamefont {T.}~\bibnamefont
  {Yamashita}}, \bibinfo {author} {\bibfnamefont {S.}~\bibnamefont {Ohara}}, \
  and\ \bibinfo {author} {\bibfnamefont {I.}~\bibnamefont {Sakamoto}},\
  }\href@noop {} {\bibfield  {journal} {\bibinfo  {journal} {J. Phys. Soc.
  Japan}\ }\textbf {\bibinfo {volume} {80}},\ \bibinfo {pages} {SA080}
  (\bibinfo {year} {2011})}\BibitemShut {NoStop}%
\bibitem [{\citenamefont {Kishine}, \citenamefont {Inoue},\ and\ \citenamefont
  {Yoshida}(2005)}]{Kishine2005}%
  \BibitemOpen
  \bibfield  {author} {\bibinfo {author} {\bibfnamefont {J.}~\bibnamefont
  {Kishine}}, \bibinfo {author} {\bibfnamefont {K.}~\bibnamefont {Inoue}}, \
  and\ \bibinfo {author} {\bibfnamefont {Y.}~\bibnamefont {Yoshida}},\
  }\href@noop {} {\bibfield  {journal} {\bibinfo  {journal} {Prog. Theor. Phys.
  Suppl.}\ }\textbf {\bibinfo {volume} {159}},\ \bibinfo {pages} {82} (\bibinfo
  {year} {2005})}\BibitemShut {NoStop}%
\bibitem [{\citenamefont {Togawa}\ \emph {et~al.}(2016)\citenamefont {Togawa},
  \citenamefont {Kousaka}, \citenamefont {Inoue},\ and\ \citenamefont
  {Kishine}}]{Togawa2016}%
  \BibitemOpen
  \bibfield  {author} {\bibinfo {author} {\bibfnamefont {Y.}~\bibnamefont
  {Togawa}}, \bibinfo {author} {\bibfnamefont {Y.}~\bibnamefont {Kousaka}},
  \bibinfo {author} {\bibfnamefont {K.}~\bibnamefont {Inoue}}, \ and\ \bibinfo
  {author} {\bibfnamefont {J.}~\bibnamefont {Kishine}},\ }\href@noop {}
  {\bibfield  {journal} {\bibinfo  {journal} {J. Phys. Soc. Japan}\ }\textbf
  {\bibinfo {volume} {85}},\ \bibinfo {pages} {112001} (\bibinfo {year}
  {2016})}\BibitemShut {NoStop}%
\bibitem [{\citenamefont {Matsumura}\ \emph {et~al.}(2017)\citenamefont
  {Matsumura}, \citenamefont {Kita}, \citenamefont {Yoshikawa}, \citenamefont
  {Michimura}, \citenamefont {Inami}, \citenamefont {Kousaka}, \citenamefont
  {Inoue},\ and\ \citenamefont {Ohara}}]{Matsumura2017}%
  \BibitemOpen
  \bibfield  {author} {\bibinfo {author} {\bibfnamefont {T.}~\bibnamefont
  {Matsumura}}, \bibinfo {author} {\bibfnamefont {Y.}~\bibnamefont {Kita}},
  \bibinfo {author} {\bibfnamefont {Y.}~\bibnamefont {Yoshikawa}}, \bibinfo
  {author} {\bibfnamefont {S.}~\bibnamefont {Michimura}}, \bibinfo {author}
  {\bibfnamefont {T.}~\bibnamefont {Inami}}, \bibinfo {author} {\bibfnamefont
  {Y.}~\bibnamefont {Kousaka}}, \bibinfo {author} {\bibfnamefont
  {K.}~\bibnamefont {Inoue}}, \ and\ \bibinfo {author} {\bibfnamefont
  {S.}~\bibnamefont {Ohara}},\ }\href@noop {} {\bibfield  {journal} {\bibinfo
  {journal} {J. Phys. Soc. Japan}\ }\textbf {\bibinfo {volume} {86}},\ \bibinfo
  {pages} {124702} (\bibinfo {year} {2017})}\BibitemShut {NoStop}%
\bibitem [{\citenamefont {Ninomiya}\ \emph {et~al.}(2018)\citenamefont
  {Ninomiya}, \citenamefont {Sato}, \citenamefont {Inoue},\ and\ \citenamefont
  {Ohara}}]{Ninomiya2018}%
  \BibitemOpen
  \bibfield  {author} {\bibinfo {author} {\bibfnamefont {H.}~\bibnamefont
  {Ninomiya}}, \bibinfo {author} {\bibfnamefont {T.}~\bibnamefont {Sato}},
  \bibinfo {author} {\bibfnamefont {K.}~\bibnamefont {Inoue}}, \ and\ \bibinfo
  {author} {\bibfnamefont {S.}~\bibnamefont {Ohara}},\ }\href@noop {}
  {\bibfield  {journal} {\bibinfo  {journal} {Physica B}\ }\textbf {\bibinfo
  {volume} {536}},\ \bibinfo {pages} {654} (\bibinfo {year}
  {2018})}\BibitemShut {NoStop}%
\bibitem [{\citenamefont {Nishikawa}\ and\ \citenamefont
  {Hukushima}(2016)}]{Nishikawa2016}%
  \BibitemOpen
  \bibfield  {author} {\bibinfo {author} {\bibfnamefont {Y.}~\bibnamefont
  {Nishikawa}}\ and\ \bibinfo {author} {\bibfnamefont {K.}~\bibnamefont
  {Hukushima}},\ }\href@noop {} {\bibfield  {journal} {\bibinfo  {journal}
  {Phys. Rev. B}\ }\textbf {\bibinfo {volume} {94}},\ \bibinfo {pages} {064428}
  (\bibinfo {year} {2016})}\BibitemShut {NoStop}%
\bibitem [{\citenamefont {Shinozaki}\ \emph {et~al.}(2016)\citenamefont
  {Shinozaki}, \citenamefont {Hoshino}, \citenamefont {Masaki}, \citenamefont
  {Kishine},\ and\ \citenamefont {Kato}}]{Shinozaki2016}%
  \BibitemOpen
  \bibfield  {author} {\bibinfo {author} {\bibfnamefont {M.}~\bibnamefont
  {Shinozaki}}, \bibinfo {author} {\bibfnamefont {S.}~\bibnamefont {Hoshino}},
  \bibinfo {author} {\bibfnamefont {Y.}~\bibnamefont {Masaki}}, \bibinfo
  {author} {\bibfnamefont {J.}~\bibnamefont {Kishine}}, \ and\ \bibinfo
  {author} {\bibfnamefont {Y.}~\bibnamefont {Kato}},\ }\href@noop {} {\bibfield
   {journal} {\bibinfo  {journal} {J. Phys. Soc. Japan}\ }\textbf {\bibinfo
  {volume} {85}},\ \bibinfo {pages} {074710} (\bibinfo {year}
  {2016})}\BibitemShut {NoStop}%
\bibitem [{\citenamefont {Laliena}, \citenamefont {Campo},\ and\ \citenamefont
  {Kousaka}(2016)}]{Laliena2016-2}%
  \BibitemOpen
  \bibfield  {author} {\bibinfo {author} {\bibfnamefont {V.}~\bibnamefont
  {Laliena}}, \bibinfo {author} {\bibfnamefont {J.}~\bibnamefont {Campo}}, \
  and\ \bibinfo {author} {\bibfnamefont {Y.}~\bibnamefont {Kousaka}},\
  }\href@noop {} {\bibfield  {journal} {\bibinfo  {journal} {Phys. Rev. B}\
  }\textbf {\bibinfo {volume} {94}},\ \bibinfo {pages} {094439} (\bibinfo
  {year} {2016})}\BibitemShut {NoStop}%
\bibitem [{\citenamefont {Masaki}\ and\ \citenamefont
  {Stamp}(2017)}]{Masaki2017}%
  \BibitemOpen
  \bibfield  {author} {\bibinfo {author} {\bibfnamefont {Y.}~\bibnamefont
  {Masaki}}\ and\ \bibinfo {author} {\bibfnamefont {R.~L.}\ \bibnamefont
  {Stamp}},\ }\href@noop {} {\bibfield  {journal} {\bibinfo  {journal} {Phys.
  Rev. B}\ }\textbf {\bibinfo {volume} {95}},\ \bibinfo {pages} {024418}
  (\bibinfo {year} {2017})}\BibitemShut {NoStop}%
\bibitem [{\citenamefont {Laliena}, \citenamefont {Campo},\ and\ \citenamefont
  {Kousaka}(2017)}]{Laliena2017}%
  \BibitemOpen
  \bibfield  {author} {\bibinfo {author} {\bibfnamefont {V.}~\bibnamefont
  {Laliena}}, \bibinfo {author} {\bibfnamefont {J.}~\bibnamefont {Campo}}, \
  and\ \bibinfo {author} {\bibfnamefont {Y.}~\bibnamefont {Kousaka}},\
  }\href@noop {} {\bibfield  {journal} {\bibinfo  {journal} {Phys. Rev. B}\
  }\textbf {\bibinfo {volume} {95}},\ \bibinfo {pages} {224410} (\bibinfo
  {year} {2017})}\BibitemShut {NoStop}%
\bibitem [{\citenamefont {Kishine}, \citenamefont {Proskurin},\ and\
  \citenamefont {Ovchinnikov}(2011)}]{Kishine2011}%
  \BibitemOpen
  \bibfield  {author} {\bibinfo {author} {\bibfnamefont {J.}~\bibnamefont
  {Kishine}}, \bibinfo {author} {\bibfnamefont {I.~V.}\ \bibnamefont
  {Proskurin}}, \ and\ \bibinfo {author} {\bibfnamefont {A.~S.}\ \bibnamefont
  {Ovchinnikov}},\ }\href@noop {} {\bibfield  {journal} {\bibinfo  {journal}
  {Phys. Rev. Lett.}\ }\textbf {\bibinfo {volume} {107}},\ \bibinfo {pages}
  {017205} (\bibinfo {year} {2011})}\BibitemShut {NoStop}%
\bibitem [{\citenamefont {Okumura}, \citenamefont {Kato},\ and\ \citenamefont
  {Motome}(2017)}]{Okumura2017}%
  \BibitemOpen
  \bibfield  {author} {\bibinfo {author} {\bibfnamefont {S.}~\bibnamefont
  {Okumura}}, \bibinfo {author} {\bibfnamefont {Y.}~\bibnamefont {Kato}}, \
  and\ \bibinfo {author} {\bibfnamefont {Y.}~\bibnamefont {Motome}},\
  }\href@noop {} {\bibfield  {journal} {\bibinfo  {journal} {J. Phys. Soc.
  Japan}\ }\textbf {\bibinfo {volume} {86}},\ \bibinfo {pages} {063701}
  (\bibinfo {year} {2017})}\BibitemShut {NoStop}%
\bibitem [{\citenamefont {Okumura}, \citenamefont {Kato},\ and\ \citenamefont
  {Motome}(2018{\natexlab{a}})}]{Okumura2018JPSJ}%
  \BibitemOpen
  \bibfield  {author} {\bibinfo {author} {\bibfnamefont {S.}~\bibnamefont
  {Okumura}}, \bibinfo {author} {\bibfnamefont {Y.}~\bibnamefont {Kato}}, \
  and\ \bibinfo {author} {\bibfnamefont {Y.}~\bibnamefont {Motome}},\
  }\href@noop {} {\bibfield  {journal} {\bibinfo  {journal} {J. Phys. Soc.
  Japan}\ }\textbf {\bibinfo {volume} {87}},\ \bibinfo {pages} {033708}
  (\bibinfo {year} {2018}{\natexlab{a}})}\BibitemShut {NoStop}%
\bibitem [{\citenamefont {Okumura}, \citenamefont {Kato},\ and\ \citenamefont
  {Motome}(2018{\natexlab{b}})}]{Okumura2018PhysB}%
  \BibitemOpen
  \bibfield  {author} {\bibinfo {author} {\bibfnamefont {S.}~\bibnamefont
  {Okumura}}, \bibinfo {author} {\bibfnamefont {Y.}~\bibnamefont {Kato}}, \
  and\ \bibinfo {author} {\bibfnamefont {Y.}~\bibnamefont {Motome}},\
  }\href@noop {} {\bibfield  {journal} {\bibinfo  {journal} {Physica B}\
  }\textbf {\bibinfo {volume} {538}},\ \bibinfo {pages} {223} (\bibinfo {year}
  {2018}{\natexlab{b}})}\BibitemShut {NoStop}%
\bibitem [{\citenamefont {Tokushuku}, \citenamefont {Kishine},\ and\
  \citenamefont {Ogata}(2017)}]{Tokushuku2017}%
  \BibitemOpen
  \bibfield  {author} {\bibinfo {author} {\bibfnamefont {K.}~\bibnamefont
  {Tokushuku}}, \bibinfo {author} {\bibfnamefont {J.}~\bibnamefont {Kishine}},
  \ and\ \bibinfo {author} {\bibfnamefont {M.}~\bibnamefont {Ogata}},\
  }\href@noop {} {\bibfield  {journal} {\bibinfo  {journal} {J. Phys. Soc.
  Japan}\ }\textbf {\bibinfo {volume} {86}},\ \bibinfo {pages} {124701}
  (\bibinfo {year} {2017})}\BibitemShut {NoStop}%
\bibitem [{\citenamefont {Watanabe}, \citenamefont {Hoshi},\ and\ \citenamefont
  {Ohe}(2016)}]{Watanabe2016}%
  \BibitemOpen
  \bibfield  {author} {\bibinfo {author} {\bibfnamefont {H.}~\bibnamefont
  {Watanabe}}, \bibinfo {author} {\bibfnamefont {K.}~\bibnamefont {Hoshi}}, \
  and\ \bibinfo {author} {\bibfnamefont {J.}~\bibnamefont {Ohe}},\ }\href@noop
  {} {\bibfield  {journal} {\bibinfo  {journal} {Phys. Rev. B}\ }\textbf
  {\bibinfo {volume} {94}},\ \bibinfo {pages} {125143} (\bibinfo {year}
  {2016})}\BibitemShut {NoStop}%
\bibitem [{SM()}]{SM}%
  \BibitemOpen
  \href@noop {} {}\bibinfo {note} {See Supplementary Material for $J$
  dependence of the nonreciprocal spin current.}\BibitemShut {Stop}%
\bibitem [{\citenamefont {Ando}(1991)}]{Ando1991}%
  \BibitemOpen
  \bibfield  {author} {\bibinfo {author} {\bibfnamefont {T.}~\bibnamefont
  {Ando}},\ }\href@noop {} {\bibfield  {journal} {\bibinfo  {journal} {Phys.
  Rev. B}\ }\textbf {\bibinfo {volume} {44}},\ \bibinfo {pages} {8017}
  (\bibinfo {year} {1991})}\BibitemShut {NoStop}%
\bibitem [{\citenamefont {Ohe}, \citenamefont {Ohtsuki},\ and\ \citenamefont
  {Kramer}(2007)}]{Ohe2007}%
  \BibitemOpen
  \bibfield  {author} {\bibinfo {author} {\bibfnamefont {J.}~\bibnamefont
  {Ohe}}, \bibinfo {author} {\bibfnamefont {T.}~\bibnamefont {Ohtsuki}}, \ and\
  \bibinfo {author} {\bibfnamefont {B.}~\bibnamefont {Kramer}},\ }\href@noop {}
  {\bibfield  {journal} {\bibinfo  {journal} {Phys. Rev. B}\ }\textbf {\bibinfo
  {volume} {75}},\ \bibinfo {pages} {245313} (\bibinfo {year}
  {2007})}\BibitemShut {NoStop}%
\bibitem [{\citenamefont {Jauho}, \citenamefont {Wingreen},\ and\ \citenamefont
  {Meir}(1994)}]{Jauho1994}%
  \BibitemOpen
  \bibfield  {author} {\bibinfo {author} {\bibfnamefont {A.-P.}\ \bibnamefont
  {Jauho}}, \bibinfo {author} {\bibfnamefont {N.~S.}\ \bibnamefont {Wingreen}},
  \ and\ \bibinfo {author} {\bibfnamefont {Y.}~\bibnamefont {Meir}},\
  }\href@noop {} {\bibfield  {journal} {\bibinfo  {journal} {Phys. Rev. B}\
  }\textbf {\bibinfo {volume} {50}},\ \bibinfo {pages} {5528} (\bibinfo {year}
  {1994})}\BibitemShut {NoStop}%
\bibitem [{Note1()}]{Note1}%
  \BibitemOpen
  \bibinfo {note} {We confirm this picture by examining the dependence on the
  strength of the spin-charge coupling $J$~\cite {SM}}\BibitemShut {NoStop}%
\bibitem [{\citenamefont {Hamamoto}\ \emph {et~al.}(2017)\citenamefont
  {Hamamoto}, \citenamefont {Ezawa}, \citenamefont {Kim}, \citenamefont
  {Morimoto},\ and\ \citenamefont {Nagaosa}}]{Hamamoto2017}%
  \BibitemOpen
  \bibfield  {author} {\bibinfo {author} {\bibfnamefont {K.}~\bibnamefont
  {Hamamoto}}, \bibinfo {author} {\bibfnamefont {M.}~\bibnamefont {Ezawa}},
  \bibinfo {author} {\bibfnamefont {K.~W.}\ \bibnamefont {Kim}}, \bibinfo
  {author} {\bibfnamefont {T.}~\bibnamefont {Morimoto}}, \ and\ \bibinfo
  {author} {\bibfnamefont {N.}~\bibnamefont {Nagaosa}},\ }\href@noop {}
  {\bibfield  {journal} {\bibinfo  {journal} {Phys. Rev. B}\ }\textbf {\bibinfo
  {volume} {95}},\ \bibinfo {pages} {224430} (\bibinfo {year}
  {2017})}\BibitemShut {NoStop}%
\bibitem [{\citenamefont {Togawa}\ \emph {et~al.}(2015)\citenamefont {Togawa},
  \citenamefont {Koyama}, \citenamefont {Nishimori}, \citenamefont {Matsumoto},
  \citenamefont {McVitie}, \citenamefont {McGrouther}, \citenamefont {Stamps},
  \citenamefont {Kousaka}, \citenamefont {Akimitsu}, \citenamefont {Nishihara},
  \citenamefont {Inoue}, \citenamefont {Bostrem}, \citenamefont {Sinitsyn},
  \citenamefont {Ovchinnikov},\ and\ \citenamefont {Kishine}}]{Togawa2015}%
  \BibitemOpen
  \bibfield  {author} {\bibinfo {author} {\bibfnamefont {Y.}~\bibnamefont
  {Togawa}}, \bibinfo {author} {\bibfnamefont {T.}~\bibnamefont {Koyama}},
  \bibinfo {author} {\bibfnamefont {Y.}~\bibnamefont {Nishimori}}, \bibinfo
  {author} {\bibfnamefont {Y.}~\bibnamefont {Matsumoto}}, \bibinfo {author}
  {\bibfnamefont {S.}~\bibnamefont {McVitie}}, \bibinfo {author} {\bibfnamefont
  {D.}~\bibnamefont {McGrouther}}, \bibinfo {author} {\bibfnamefont {R.~L.}\
  \bibnamefont {Stamps}}, \bibinfo {author} {\bibfnamefont {Y.}~\bibnamefont
  {Kousaka}}, \bibinfo {author} {\bibfnamefont {J.}~\bibnamefont {Akimitsu}},
  \bibinfo {author} {\bibfnamefont {S.}~\bibnamefont {Nishihara}}, \bibinfo
  {author} {\bibfnamefont {K.}~\bibnamefont {Inoue}}, \bibinfo {author}
  {\bibfnamefont {I.~G.}\ \bibnamefont {Bostrem}}, \bibinfo {author}
  {\bibfnamefont {V.~E.}\ \bibnamefont {Sinitsyn}}, \bibinfo {author}
  {\bibfnamefont {A.~S.}\ \bibnamefont {Ovchinnikov}}, \ and\ \bibinfo {author}
  {\bibfnamefont {J.}~\bibnamefont {Kishine}},\ }\href@noop {} {\bibfield
  {journal} {\bibinfo  {journal} {Phys. Rev. B}\ }\textbf {\bibinfo {volume}
  {92}},\ \bibinfo {pages} {220412} (\bibinfo {year} {2015})}\BibitemShut
  {NoStop}%
\bibitem [{\citenamefont {Chepiga}\ \emph {et~al.}(2017)\citenamefont
  {Chepiga}, \citenamefont {Ki}, \citenamefont {Li}, \citenamefont {Li},
  \citenamefont {Zhu}, \citenamefont {Kato}, \citenamefont {Ovchinnikova},
  \citenamefont {Mila}, \citenamefont {Martin}, \citenamefont {Mandrus},\ and\
  \citenamefont {Morpurgo}}]{Wang2017}%
  \BibitemOpen
  \bibfield  {author} {\bibinfo {author} {\bibfnamefont {N.}~\bibnamefont
  {Chepiga}}, \bibinfo {author} {\bibfnamefont {D.-K.}\ \bibnamefont {Ki}},
  \bibinfo {author} {\bibfnamefont {L.}~\bibnamefont {Li}}, \bibinfo {author}
  {\bibfnamefont {F.}~\bibnamefont {Li}}, \bibinfo {author} {\bibfnamefont
  {W.}~\bibnamefont {Zhu}}, \bibinfo {author} {\bibfnamefont {Y.}~\bibnamefont
  {Kato}}, \bibinfo {author} {\bibfnamefont {O.~S.}\ \bibnamefont
  {Ovchinnikova}}, \bibinfo {author} {\bibfnamefont {F.}~\bibnamefont {Mila}},
  \bibinfo {author} {\bibfnamefont {I.}~\bibnamefont {Martin}}, \bibinfo
  {author} {\bibfnamefont {D.}~\bibnamefont {Mandrus}}, \ and\ \bibinfo
  {author} {\bibfnamefont {A.~F.}\ \bibnamefont {Morpurgo}},\ }\href@noop {}
  {\bibfield  {journal} {\bibinfo  {journal} {Phys. Rev. Lett.}\ }\textbf
  {\bibinfo {volume} {118}},\ \bibinfo {pages} {257203} (\bibinfo {year}
  {2017})}\BibitemShut {NoStop}%
\bibitem [{\citenamefont {Aoki}, \citenamefont {Togawa},\ and\ \citenamefont
  {Ohara}(2018)}]{Aoki2018}%
  \BibitemOpen
  \bibfield  {author} {\bibinfo {author} {\bibfnamefont {R.}~\bibnamefont
  {Aoki}}, \bibinfo {author} {\bibfnamefont {Y.}~\bibnamefont {Togawa}}, \ and\
  \bibinfo {author} {\bibfnamefont {S.}~\bibnamefont {Ohara}},\ }\href@noop {}
  {\bibfield  {journal} {\bibinfo  {journal} {Phys. Rev. B}\ }\textbf {\bibinfo
  {volume} {97}},\ \bibinfo {pages} {214414} (\bibinfo {year}
  {2018})}\BibitemShut {NoStop}%
\bibitem [{\citenamefont {Bostrem}, \citenamefont {Kishine},\ and\
  \citenamefont {Ovchinnikov}(2008{\natexlab{a}})}]{Bostrem2008-1}%
  \BibitemOpen
  \bibfield  {author} {\bibinfo {author} {\bibfnamefont {I.~G.}\ \bibnamefont
  {Bostrem}}, \bibinfo {author} {\bibfnamefont {J.}~\bibnamefont {Kishine}}, \
  and\ \bibinfo {author} {\bibfnamefont {A.~S.}\ \bibnamefont {Ovchinnikov}},\
  }\href@noop {} {\bibfield  {journal} {\bibinfo  {journal} {Phys. Rev. B}\
  }\textbf {\bibinfo {volume} {77}},\ \bibinfo {pages} {132405} (\bibinfo
  {year} {2008}{\natexlab{a}})}\BibitemShut {NoStop}%
\bibitem [{\citenamefont {Bostrem}, \citenamefont {Kishine},\ and\
  \citenamefont {Ovchinnikov}(2008{\natexlab{b}})}]{Bostrem2008-2}%
  \BibitemOpen
  \bibfield  {author} {\bibinfo {author} {\bibfnamefont {I.~G.}\ \bibnamefont
  {Bostrem}}, \bibinfo {author} {\bibfnamefont {J.}~\bibnamefont {Kishine}}, \
  and\ \bibinfo {author} {\bibfnamefont {A.~S.}\ \bibnamefont {Ovchinnikov}},\
  }\href@noop {} {\bibfield  {journal} {\bibinfo  {journal} {Phys. Rev. B}\
  }\textbf {\bibinfo {volume} {78}},\ \bibinfo {pages} {064425} (\bibinfo
  {year} {2008}{\natexlab{b}})}\BibitemShut {NoStop}%
\bibitem [{\citenamefont {Bostrem}\ \emph {et~al.}(2009)\citenamefont
  {Bostrem}, \citenamefont {Kishine}, \citenamefont {Lavrov},\ and\
  \citenamefont {Ovchinnikov}}]{Bostrem2009}%
  \BibitemOpen
  \bibfield  {author} {\bibinfo {author} {\bibfnamefont {I.~G.}\ \bibnamefont
  {Bostrem}}, \bibinfo {author} {\bibfnamefont {J.}~\bibnamefont {Kishine}},
  \bibinfo {author} {\bibfnamefont {R.~V.}\ \bibnamefont {Lavrov}}, \ and\
  \bibinfo {author} {\bibfnamefont {A.~S.}\ \bibnamefont {Ovchinnikov}},\
  }\href@noop {} {\bibfield  {journal} {\bibinfo  {journal} {Phys. Lett. A}\
  }\textbf {\bibinfo {volume} {373}},\ \bibinfo {pages} {558} (\bibinfo {year}
  {2009})}\BibitemShut {NoStop}%
\bibitem [{\citenamefont {Kishine}, \citenamefont {Ovchinnikov},\ and\
  \citenamefont {Proskurin}(2010)}]{Kishine2010}%
  \BibitemOpen
  \bibfield  {author} {\bibinfo {author} {\bibfnamefont {J.}~\bibnamefont
  {Kishine}}, \bibinfo {author} {\bibfnamefont {A.~S.}\ \bibnamefont
  {Ovchinnikov}}, \ and\ \bibinfo {author} {\bibfnamefont {I.~V.}\ \bibnamefont
  {Proskurin}},\ }\href@noop {} {\bibfield  {journal} {\bibinfo  {journal}
  {Phys. Rev. B}\ }\textbf {\bibinfo {volume} {82}},\ \bibinfo {pages} {064407}
  (\bibinfo {year} {2010})}\BibitemShut {NoStop}%
\end{thebibliography}%

\end{document}